\RequirePackage{fix-cm}
\documentclass[]{svjour3}                     % onecolumn (standard format)
\smartqed  % flush right qed marks, e.g. at end of proof
\usepackage{pdflscape}
\usepackage{url}
\usepackage{hyperref}
\usepackage{booktabs}
\usepackage{graphicx}
\usepackage{amsmath}
\usepackage{multirow}
\usepackage{tabularx}
\usepackage{amsfonts}
\usepackage{amssymb}
\usepackage{pgfplots}
\usepackage{tikz}
\usepackage{multirow}
\usepackage{epstopdf}
\usetikzlibrary{arrows}
\usepackage{pgfplots}
\usepackage{textcomp}
\usepackage[utf8]{inputenc}
\usepackage{siunitx}
\usepackage[section]{placeins}
\usepackage{caption}

\usepackage{algorithm}
\usepackage{ amssymb }
\usepackage{algpseudocode}
\usepackage[numbers]{natbib}
\journalname{Journal of Latex Class}
\begin{document}
	
	\title{Machine Learning based Intrusion Detection Systems for IoT Applications}
	
	\titlerunning{Intrusion Detection Systems for IoT Applications}        % if too long for running head
	
	\author{Abhishek Verma$^{1,*}$\and
		Virender Ranga$^{1}$
	}

	\institute{$^{1}$Department of Computer Engineering, National Institute of Technology, Kurukshetra, India \\
		\email{$^{*}$abhiverma866@gmail.com} \\ 
		\email{virender.ranga@nitkkr.ac.in} \\         %  \\
	}
	
	\date{Received: 8 August 2019 / Accepted: date}
	% The correct dates will be entered by the editor
	\maketitle
\begin{abstract}
Internet of Things (IoT) and its applications are the most popular research areas at present. The characteristics of IoT on one side make it easily applicable to real-life applications, whereas on the other side expose it to cyber threats. Denial of Service (DoS) is one of the most catastrophic attacks against IoT. In this paper, we investigate the prospects of using machine learning classification algorithms for securing IoT against DoS attacks. \textcolor{black}{A comprehensive study is carried on the classifiers which can advance the development of anomaly-based intrusion detection systems (IDSs).} Performance assessment of classifiers is done in terms of prominent metrics and validation methods. Popular datasets CIDDS-001, UNSW-NB15, and NSL-KDD are used for benchmarking classifiers. Friedman and Nemenyi tests are employed to analyze the significant differences among classifiers statistically. In addition, Raspberry Pi is used to evaluate the response time of classifiers on \textcolor{black}{IoT specific hardware}. We also discuss a methodology for selecting the best classifier as per application requirements. The main goals of this study are to motivate IoT security researchers for developing IDSs using ensemble learning, and suggesting appropriate methods for statistical assessment of classifier's performance.\footnote{The final publication is
	available at https://link.springer.com/article/10.1007/s11277-019-06986-8}
\end{abstract}


\section{Introduction}\label{Introduction}
\textcolor{black}{Security and privacy aspects of the Internet of Things (IoT) \cite{ashton2009internet, IoTEnablingTechnologies, Verma2019WPC,verma2019elnids, medhat} are the key players which drive its potential to become one of the globally adopted technology in the future \cite{RIAHISFAR2018118, Granjal}.} However, self-configuring and open nature of IoT makes it vulnerable to various insider and outsider attackers \cite{mosenia2017comprehensive, Hwang:2015}. \textcolor{black}{Attackers may compromise the users' security and privacy in order to gain access to their personal information, create monetary losses, and eavesdropping \cite{ziegeldorf2014privacy}.} These factors prevent global adoption of IoT, consequently slowing down its growth \cite{gao2014unified}. \textcolor{black}{Denial of Service (DoS) is one of the most catastrophic attacks that prevent legitimate user to access the service it has paid for \cite{roman2013features, das}}. This violates Service Level Agreement (SLA) terms which leads to huge monetary losses for firms and organizations. Moreover, DoS also affects the services of small networks, i.e., smart home, healthcare and smart agriculture etc \cite{li2011smart}. DoS attacks on critical smart applications such as healthcare may even lead to death like situations because normal services get delayed \cite{moosavi2015sea}. \textcolor{black}{IoT devices (e.g., smart light bulbs, smart door locks, smart television) are an easy target of attackers which exploit their vulnerabilities in order to perform DoS attacks \cite{notra2014experimental,ronen2016extended, sivaraman2015network, dhanjani2013hacking, zahoor}}. Thus, securing these devices is one of the important concerns for researchers nowadays \cite{zhao2013survey,arics2015internet}. \textcolor{black}{To address this issue, intrusion detection is being heavily researched worldwide \cite{Zarpelao2017, baykara}.} Intrusion detection systems (IDSs) are categorized into three classes based on the detection method, i.e., signature, anomaly, and specification. Among three IDS types, our focus is primarily on anomaly-based IDS \cite{garcia2009anomaly}. A signature-based IDS matches network traffic patterns with the attack patterns (signatures) already stored in its database. In case a match is found, an alarm is raised. A signature-based IDS has high accuracy and low false alarm rate, however, it is incapable of detecting new attacks. A specification-based network IDS match traffic behavior (parameters) against a predefined set of rules and values (specifications) for detecting malicious activities. These specifications are manually specified by a security expert. In contrast to signature and specification based IDS, anomaly-based IDS continuously checks network traffic for any deviation from normal network profile. \textcolor{black}{In case a deviation exceeds the threshold, an alarm is raised to signify attack detection.} The normal network profile is learned using machine learning (ML) algorithms. A anomaly-based IDS are preferred over signature and specification based IDS because of its ability to detect new attacks, but this comes with a cost of high false alarm rate. The effectiveness of anomaly-based IDS depends on the goodness of detection engine (model or classifier), and this goodness comes with the quality of network traffic patterns (dataset instances) being used for engine's training. Once the detection engine has been trained, it can detect new attacks effectively. Intrusion detection in IoT networks is characterized as a binary classification problem in which a trained classifier aims to classify network traffic into normal or attack class with maximum accuracy and minimum false alarms (FAR). The high performance of the classifier in terms of accuracy and FAR solely depends on the choice of classification algorithm and training data. Security experts prefer good performing classifier for the task of intrusion detection. Many solutions for intrusion detection have been proposed in the literature \cite{debar2000revised, axelsson2000intrusion, modi2013survey, lunt1993survey, garcia2009anomaly, butun2014survey} and most of them are dedicated for traditional networking paradigms only, and as far as the literature is concerned, less work has been done towards the development of ML-based intrusion detection for IoT applications. Moreover, we didn't find any work in the literature which statistically analyzed the significance of classifier's performance in particular to IoT based intrusion detection. Also, no work in the literature has realized the execution of classifier on an IoT hardware. Thus, our focus is basically on utilizing ML classification algorithms for building IDS in order to secure IoT against DoS attacks.  

In this work, we carry out a performance assessment of ML classifiers for IDS in particular to IoT. The performance of single classifiers including CART and MLP, and classifier ensembles namely Random forest (RF), AdaBoost (AB), Extreme gradient boosting (XGB), Gradient boosted machine (GBM), and Extremely randomized trees (ETC) is measured in terms of prominent metrics, i.e., accuracy, specificity, sensitivity, false positive rate, area under the receiver operating characteristic curve (AUC).  Hyper-tuning (finding the set of optimal parameters) of all the classifiers is done using random search\cite{randomsearch}. The significant differences of classifiers are statistically assessed using a well known statistical test. \textcolor{black}{Finally, we have tested the performance of classifiers in terms of average response time on Raspberry Pi, i.e., IoT device \cite{zhao}}.    

Our primary contributions can be summarized as follows.                               

\begin{itemize}
	\item \textcolor{black}{Performance assessment of different ML classifiers on CIDDS-001, UNSW-NB15, NSL-KDD datasets with repeated hold-out and repeated cross fold validation methods is done.}   
	\item \textcolor{black}{Statistical assessment of performance results using widely used Friedman test (non-parametric statistical test) and Nemenyi post-hoc test, i.e., Friedman test for classifier significance test and Nemenyi test for pairwise comparison among classifiers is done.}  
	\item \textcolor{black}{Implementation and execution of classifiers on Raspberry Pi hardware for realizing actual response time on real IoT hardware is carried out.}   
\end{itemize}

\textcolor{black}{The paper organization is as follows. Section \ref{RelatedWork} discusses recent works in the concerned domain. Section \ref{ClassificationAlgorithms} provides a brief discussion on classification algorithms, i.e., single classifiers, and ensembles. Experimental design is discussed in section \ref{ExperimentalDesign}.}  Discussion related to the classifier's performance and statistical tests is done in section \ref{ResultsandAnalysis}. Section \ref{Conclusion} concludes the paper.   
\section{Related Work}\label{RelatedWork}
There are few works present in the literature which suggest methods for defending IoT against DoS attacks. Like, Misra \textit{\textit{et al.}} \cite{misra2011learning} proposed a specification-based IDS based on Learning Automata for preventing distributed DoS attacks against IoT. The authors considered preventing IoT middle-ware layer rather than a particular device. The proposed security system sets a threshold for the number of requests a middleware layer can service. As soon as the number of incoming requests exceeds the set threshold, an attack is detected. Kasinathan \textit{et al.} \cite{kasinathan2013denial} proposed a signature-based IDS framework for detecting DoS attack in IoT. The proposed framework consists of a monitoring and detection modules. \textcolor{black}{These modules are integrated with the network framework of European Union (EU) FP7 project `ebbits' for securing the network against DoS attacks.} A DoS protection manager and IDS are integrated with the `ebbits' network. A network-based IDS is used for capturing and analyzing the packets sniffed from IDS probe node's that are spread across the network. The evaluation results show that the proposed framework performs well in terms of true positive and false positive rate.  \textcolor{black}{Kasinathan \textit{et al.} \cite{kasinathan2013ids} proposed an IDS to detect DoS attacks.} Suricata \cite{suricata2014open} (open source IDS) is used for pattern matching and attack detection. A probe node is used to sniff all the packet transmissions in the network and transfer information to IDS for further analysis. Penetration testing tool `Scapy' is used to test the performance of the proposed IDS. No simulation study is done in support of IDS performance and its usability. \textcolor{black}{Moreover, the authors did not mention any details regarding signature database management (update).} Lee \textit{et al.} \cite{lee2014lightweight} proposed a novel IDS for detecting DoS attacks. The key idea behind the proposed IDS is to analyze the node's energy consumption in order to track malicious nodes. \textcolor{black}{The authors proposed various models for normal energy consumption in mesh routing based networks.} The proposed security system requires nodes to monitor their own energy consumption at a sampling rate of 0.5 seconds. The proposed IDS continuously checks the energy consumption of nodes against the defined threshold, and whenever a deviation is found for any node, such a node is marked as malicious and removed from the routing table. The proposed approach shows promising results in terms of accuracy only. \textcolor{black}{The major concern with this proposed approach is that there is no inbuilt mechanism to verify the integrity of energy consumption values being reported by a node.} Sonar \textit{et al.} \cite{sonar2016approach} proposed an IDS to detect distributed DoS attacks in IoT networks. The authors implemented IDS as software-based manager deployed between network and gateway. The proposed IDS maintains greylist and blacklist or IP addresses in order to control the access to the network. \textcolor{black}{In the proposed IDS, the greylist is updated every 40 seconds while blacklist is updated every 300 seconds.} Simulation of the proposed IDS is performed on the contiki \cite{dunkels2004contiki} operating system. The proposed IDS performs do not achieve satisfactory performance in terms of packet delivery ratio, the number of serviced packets, true positives, and false positives. Moreover, the recovery time is larger than agent learning time which adds to the major limitations of the proposed IDS. Diro \textit{et al.} \cite{diro2018distributed} proposed a deep learning based IDS  for defending DoS against IoT networks. The proposed model is evaluated using NSL-KDD dataset. The authors performed the comparison of proposed IDS with the traditional shallow model approach. In addition, the proposed IDS is implemented with centralized and distributed detection scheme. The comparison results show that the distributed attack detection scheme performs better compared to centralized detection scheme in terms of accuracy. Similarly, the deep model shows better results compared to the shallow model in terms of accuracy, precision, recall, and F1 measure. \textcolor{black}{Tama \textit{et al.} \cite{tama2019depth} proposed an anomaly based IDS that uses gradient boosted machine (GBM) as a detection engine. \textcolor{black}{The optimal parameters of GBM are obtained using grid search and the performance of the proposed IDS is validated using hold-out and cross fold methods on three different datasets namely UNSW-NB15, NSL-KDD, and GPRS. The authors show that proposed IDS outperforms the fuzzy classifier, GAR forest, tree based ensembles in terms of accuracy, specificity, sensitivity, and area under curve (AUC).} Primartha \textit{et al.} \cite{primartha2017anomaly} studied the performance of RF based IDS in terms of accuracy and false alarm rate. The authors employed NSL-KDD, UNSW-NB15 and GPRS dataset for model training and testing. \textcolor{black}{The proposed IDS is studied with different tree size ensembles, and statistical analysis based on Friedman ranking showed that the ensemble of 800 trees achieves best results whereas an ensemble of 20 trees showed the worst performance.} Moreover, the proposed RF based IDS outperforms ensemble of Random tree$ + $Naive Bayes, and single classifiers like NBTree and Multi-layer perceptron.}

\section{Classification Algorithms}\label{ClassificationAlgorithms}
Wolpert \textit{et al.} \cite{wolpert1997no} stated a theorem popularly known as ``no free lunch” theorem that shows the importance of experimenting with different machine classifiers for solving classification tasks. The theorem states that ``there is no single learning algorithm that universally performs best across all domains" \cite{douglas2011performance}. Thus, different classifiers should be tested for solving domain specific problems, and in our case, the problem is intrusion detection or classification problem. We consider two types of classification algorithms, i.e., ensembles and single classifiers. Among ensembles, widely studied algorithms \cite{galar2011review,krawczyk2017ensemble,sagi} like Random forest (RF), AdaBoost (AB), Gradient boosted machine (GBM), Extreme gradient boosting (ETC), and Extremely randomized trees (ETC) are chosen. There are main reasons for selection of mentioned classification algorithms. First, because ensemble-based classification methods are prone to over-fitting in case the number of input features is large we also choose to study some single classifiers like Classification and regression trees (CART), and Multi-layer perceptron (MLP). Second, the performance of ensembles has not been studied in depth for CIDDS-001 and UNSW-NB15 datasets. \textcolor{black}{Third, the performance of ensembles and single classifiers over real IoT hardware has not been studied yet, which motivated us for carrying this analytical study.}

\subsection{Classifier ensembles}
This section discusses various classifier ensembles in brief. Ensembles have been proven to be good classification and regression algorithms in the literature. \textcolor{black}{Thus, we have used five different ensembles in this analytical study. }   
\subsubsection{Random forest (RF)}
RF\cite{RandomForest} is a collection of trees, i.e., predictors \{$ t(x_{in}, \theta_n), n = 1, \ldots $\} which individually make predictions on a given input $ x_{in} $. Each predictor depends on the random set of variables \{$ \theta_n $\} that are sampled independently with the same distribution. \textcolor{black}{The main idea behind RF is that the number of predictors together might achieve better prediction accuracy while avoiding the over-fitting problem.} Each predictor in RF grows to a maximum size without getting pruned. Once a large number of trees are created, they make predictions over the input data by voting for the most popular class at input $ x_{in} $. \textcolor{black}{For the performance assessment the  number of estimators (trees) is set to $ 500 $ and maximum depth for tree construction is set to $ 26 $ as recommended by \cite{RandomForest,tama2019depth}. The other parameters are obtained using randomized search.}  

%In this study, random search \cite{randomsearch} is employed for finding the optimal set of RF parameters i.e., hyper-tuning.

\subsubsection{AdaBoost (AB)}
AB\cite{AdaBoost} is an adaptive meta-estimator that learns the initial training weights on the original dataset. These weights act as input to additional copies of the classifier based on incorrectly classified instances. The subsequent classifiers adjust the weights of classified instances, i.e., difficult cases. In this way, AB improves the performance of learning algorithms by boosting weak learners such that final model converges to a strong learner. Eq. \eqref{Eq1} represents a boost classifier where $ c_p $ resembles a weak learner and $ x $ resembles an input object. 

\begin{equation}
\label{Eq1}
C_P(x) = \sum_{p = 1}^{P} c_p (x)
\end{equation}

$ c_p (x) $ returns the value indicating predicted class. Each $ c_p $ generates an output hypothesis ($ h(x_i) $) for each instance in the training set. At each iteration \textit{p}, a $ c_p $ is chosen and assigned a coefficient $ \beta_p $ such that the sum training error $ E_t $ (represented as eq. \eqref{Eq2}) of the resulting $ p $-stage $ C_P(x) $ is minimized. $ C_{p-1}(x_i) $ represents the boosted classifier built from previous training phase, $ E (C) $ is error function which is to be minimized, and $ c_p(x) = \beta_p h(x_i) $ is the weak learner that is to be added to the final model, i.e., classifier. \textcolor{black}{The optimal parameters of AB include 50 estimators and 0.1 learning rate.} 

%Hyper-tuning of AB is done using random search.     

\begin{equation}
\label{Eq2}
E_t = \sum_{i}^{}E[C_{p-1}(x_i) + \beta_p h(x_i)] 
\end{equation}

\subsubsection{Gradient boosted machine (GBM)}
GBM\cite{friedman2002stochastic, friedman2001greedy} is a member of the ensemble family which aims to improve the performance of decision trees (DT). Like other boosting methods it sequentially combines weak classifiers, i.e., DT, and allows them to optimize an arbitrary differential loss function in order to form a strong prediction model. Each present learner (tree) relies on the predictions of previous learners in order to improve the prediction errors. Formally, let us consider a set of random input variables and random output represented by $ x $ (eq. \eqref{Eq3}) and $ z $ respectively.  
\begin{equation}
\label{Eq3}
x = \{x_1, x_2, \ldots, x_N\}
\end{equation}
Our aim is to find an estimate $ A $ (approximation) that maps $ x $ to $ z $ by using training data $ \{z, x_i\}^N_1 $. $ A $ is represented as Equation \ref{Eq6}. 
Given a dataset ($ S $) with $ p $ samples and $ q $ features as represented by eq. \eqref{Eq4}. Then, a ensemble utilizes M additive function to predict final output eq. \eqref{Eq5}.

\begin{equation}
\label{Eq4}
S = \{(x_i, z_i)\} (|D| = p, x_i \in \mathbb{R}^q, z_i \in \mathbb{R})
\end{equation}   

\begin{equation}
\label{Eq5}
\hat{z_i} = \phi (x_i) = \sum_{m = 1}^{M}f_m(x_i), f_k \in A
\end{equation}

\begin{equation}
\label{Eq6}
A  = \{f(x) = w_{r(x)}(r: \mathbb{R}^m \to K, w \in \mathbb{R}^K) \}
\end{equation}

$ A $ is the instance set of DT, and $ r $ represents the tree structure that relates an instance to the correlating leaf index, $ K $ indicates the total count of trees, and $ f_m $ is a single tree with structure $ r $ and leaf weight $ w $. The tree ensemble makes the final prediction by summation of the scores ($ w $) of leaves which are found by classifying given test sample. Suppose, a first classifier (i.e., tree) makes prediction $ h_1(x) $ over a sample $ \{(x_i, y_i)\}^N_1 $. Then, $ h_1(x) $ is fed as input to next classifier in order to adjust the weights of previously misclassified instances. Consequently, next classifier makes prediction $ h_2(x) $ over $ \{(x_i, y_i - h_1(x_i))\}^N_1 $.  The final prediction $ h(x) $ for a given sample data $ S $ is the summation of predictions made by the trees while minimizing the prediction error. \textcolor{black}{The hyper-tuned parameters for GBM are:  500 estimators, maximum tree construction depth is 3, minimum samples required for split are 100 and 0.1 learning rate.}

\subsubsection{Extreme gradient boosting (XGB)}
XGB\cite{xgboost} is also known as regularized gradient boosting is an improved version of GBM. Like GBM, XGB follows the same principle of gradient boosting. The only key difference between them is in terms of modeling details. XGB uses more regularized model formalization in order to control over-fitting and increase generalization ability while GBM focuses only on the variance. Regularization parameter ($ \zeta $) is mathematically expressed as eq. \eqref{Eq7}. Where, $ T_l $ is the number of leaves in the tree, $ w^2_j $ is the score on the $ j $-th leaf, $\lambda$ represents regularization term the controls model complexity. XGB uses gradient boosting for optimizing the loss function during model training. Typically, for binary classification the LogLoss function ($ L $)  \cite{Bishop:2006:PRM:1162264} is used and represented as eq. \eqref{Eq8}. Where, $ N $ is the total number of observations, $ y_i $ is the binary indicator of whether predicted class $ c $ is the correct classification for particular observation $ o $ and $ p_i $ is the predicted probability that particular observation $ o $ is of class $ c $. Most importantly, $ L $ controls the predictive power, and $ \zeta $ controls the simplicity of the model. The major implementation enhancement of XGB includes usage of sparse matrices (DMatrix) with sparsity aware algorithms, improved data structures, parallelization support. Thus, XGB leverages the hardware to achieve high speed computing with low memory utilization (primary memory and cache). \textcolor{black}{The optimal values of parameters obtained for XGB are: 100 estimators, maximum tree depth is 8, value of minimum child weight is 1, gbtree booster is considered, minimum loss reduction and sub-sample ratio are 2 and 0.6 respectively.   }

\begin{equation}
\label{Eq7}
\zeta = \gamma T_l + \frac{1}{2} \lambda \sum_{j=1}^{T} w^2_j
\end{equation}

\begin{equation}
\label{Eq8}
L = -\frac{1}{N} \sum_{i=1}^{N}(y_i log (p_i) + (1-y_i) log (1 - p_i))
\end{equation}

\subsubsection{Extremely randomized trees (ETC)}
ETC\cite{extraTrees} also known as Extra trees is a tree induction algorithm for performing supervised classification and regression. To be more specific, ETC builds an ensemble of unpruned DTs. The key procedure in ETC involves randomly selecting both features and cut-point irrespective of the target variable. At each tree node, this procedure is followed with totally or partially selecting a certain number of features among which the optimal one is determined. In the worst case, the algorithm selects a single feature and cut-point at each node. In this manner, totally randomized trees are built which are independent of the training sample's target attribute values. The classical top-down methodology is followed while building the ensemble. Unlike other tree-based ensemble algorithms, ETC uses complete training sample rather than bootstrap replicas in order to grow the trees while minimizing bias and variance. The ETC splitting procedure for numeric features has three important parameters. The first parameter is $ K $ indicates the number of features selected at each node. The second parameter is $ n_{min} $ which represents minimum training set size for splitting a node. The third parameter is $ T_{count} $ that is the number of trees in the ensemble. All three parameters play a significant role in the ETC building. Where $ K $ is responsible for the strength of feature selection procedure, $ n_{min} $ governs the averaging output noise, and $ T_{count} $ specifies the reduction in variance. ETC performs almost the same as RF, however with optimal feature selection ETC is computationally faster compared to RF. \textcolor{black}{The hypertuned parameters obtained from randomized search for ETC are: 1788 estimators, maximum tree depth value is 10, minimum sample size for split is 5, number of features considered for best split are  $ \log_{10}2 $, gini criterion is considered with no bootstrapping.}

\subsection{Single classifiers}
This section discusses single classifiers in brief. In this comparative study we have used classification and regression trees, and multi-layer perceptron. 
\subsubsection{Classification and regression trees (CART)}
CART\cite{breiman2017classification} is one of the widely employed ML methods for predictive modeling problems. It is a non-parametric algorithm with a built-in mechanism to handle missing feature values. \textcolor{black}{CART involves recursive partitioning of training samples and fitting of a simple prediction model within each partition.} This partitioning can be represented as DT graphically. CART employs exhaustive search technique, in order to identify the splitting variables such that the total impurity of node's child nodes (two children) is minimized. CART uses the Gini index as its impurity function which makes it computationally efficient over entropy based classification tree algorithms. \textcolor{black}{The number of folds in  the internal cross-validation and the minimal number of observations at the terminal nodes considered are 5 and 2, respectively as used in \cite{tama2019depth}. The optimal value of maximum depth of tree construction obtained is 10.}

\subsubsection{Multi-layer perceptron (MLP)}
\textcolor{black}{MLP\cite{haykin1994neural} is a logical unit of connected nodes (artificial neurons) that attempts to mimic the biological brain behavior commonly referred as a feed-forward artificial neural network.} It learns its expertise towards a particular task using supervised learning approaches. MLP comprises several layers, i.e., input, middle and output. \textcolor{black}{Training of MLP involves learning a mathematical function  $ f(.) $ shown in eq. \eqref{Eq9}, where $ d $, $ c $ are the number of inputs and outputs respectively.} In order to perform any predictive task, MLP learns a non-linear function approximator over a set of input features  \{$ I  = i_1, i_2, \ldots, i_d $\} and output variable $ O $, i.e., class. The leftmost layer also termed as input layer consists of many artificial neurons \{ $ i_p|i_1,i_2, \ldots,i_d $\} each representing a particular input feature. The second layer, i.e., the middle layer performs the task of transformation. First, the outputs from the former layer are summed using weighted linear summation $ y $ represented as eq.\eqref{Eq10}. Second, a non-linear activation function ($ g(\cdot) $) is applied to $ y $ which results into a value that is forwarded to further layers, typically output layer in case single hidden layer is present. The rightmost layer is the output layer which receives values from the last hidden layer and responsible for firing outputs, i.e., final predictions. \textcolor{black}{The optimal parameters values of MLP obtained are: hidden layer size of 100, logistic activation function, \textit{sgd} solver, learning rate of 0.001, and 200 maximum iterations.}

\begin{equation}
\label{Eq9}
f(\cdot) = R^d \to R^c 
\end{equation}

\begin{equation}
\label{Eq10}
y = \sum_{p = 1}^{d}(w_pi_p)
\end{equation}       

\section{Experimental Design}\label{ExperimentalDesign}
\subsection{Experimental setup}
The performance assessment has been carried out on a machine operated on 64-bit Windows 10 Pro and equipped with Intel\textsuperscript{\textregistered} i7-7700 four core CPU having 3.60 GHz clock speed and 12GB main memory. The classifiers are implemented in the Python programming language (version 3.6.1). Parameter hyper-tuning is performed on PARAM Shavak system operated on 64-bit Ubuntu 14.04 and equipped with  Intel\textsuperscript{\textregistered} Xeon\textsuperscript{\textregistered} Gold 6132 twenty eight-core CPU having 2.6GHz clock speed and 96GB main memory. Raspberry Pi 3 Model B+ operated on Raspbian operating system and equipped with 64-bit quad-core ARM CPU running having 1.4GHz clock speed and 1GB main memory is used for assessing the response time of classifiers. Popular ML library scikit-learn \cite{scikit-learn} is utilized for implementing classifiers. In order to perform statistical tests on the performance results, we used the STAC \cite{STAC} web platform application.                   

\subsection{Datasets}

In this study three different datasets, i.e., CIDDS-001 \cite{CIDDS-001}, UNSW-NB15 \cite{UNSW-NB15}, and NSL-KDD \cite{NSL-KDD} are used. \textcolor{black}{We choose CIDDS-001 and UNSW-NB15 dataset as they are most recently generated datasets and contain traffic of real data, and hence can be beneficial for building accurate IDSs for monitoring and detection of new type of DoS attacks in IoT networks.} The CIDDS-001 dataset has recently been released for facilitating the development of anomaly-based IDS. The complete dataset contains approximately 32 million records comprising of normal and attacks traffic. CIDDS-001 possesses 12 features and 2 labeling attributes. Random sampling is employed to extract 100,000 instances from the internal server traffic data, compromising of 80,000 normal and 20,000 attacks (DoS) records. The extracted sample is used for carrying out hold-out and cross fold validation tests of classifiers. Our previous works \cite{verma2018evaluation, verma2018statistical} focused on evaluating the performance of various ML classification algorithms on CIDDS-001 dataset.   

\textcolor{black}{Further, we have conducted our experiments on newly publicly available dataset known as UNSW-NB15.} The dataset possesses 49 features and 1 class attribute. A part of the dataset is used as train and test sets, i.e., UNSW\_NB15\_Train and UNSW\_NB15\_Test. The train set comprises of 175,341 instances, and the test set comprises of 82,332 instances. The train set includes 56,000 instances of normal traffic and 119,341 instances of attack traffic. Similarly, the test set includes 37,000 instances of normal traffic and 45,332 instances of attack traffic. Hold-out validation is conducted using the complete train and test sets, whereas for cross-fold validation test only train set is employed.

Subsequently, NSL-KDD dataset is also used for performing validation of classifiers. The dataset contains 41 features and 1 class attribute. In this study, KDDTrain+ (training) and KDDTest+ (testing) sets of  NSL\_KDD dataset are used. The KDDTrain+ set contains total 25,192 instances comprising of 13,499 instances of attack traffic and 11,743 instances of normal traffic. Whereas, the KDDTest+ set contains total 22,544 instances comprising of 9,711 instances of attack traffic and 12,833 instances of normal traffic. Hold-out and cross fold validation of classifiers is done on each dataset individually. The choice of these sets is done in order to avoid random sampling of instances from complete NSL-KDD dataset.                

\subsection{Evaluation metrics and Validation methods}
Selection of input parameter settings influences the overall performance of the classifiers, thus we follow random search \cite{randomsearch} procedure to find the optimal input parameters of RF, AB, XGB, GBM, and ETC for different datasets. RandomizedSearchCV implementation in \textit{scikit-learn} package of Python programming language is used for hyper-tuning of parameters. RandomizedSearchCV finds optimal parameter settings by performing a cross-validated search over candidate parameter values provided by the user. Prominent metrics for evaluating classifier's performance have been used in this study. These metrics include accuracy, specificity or true negative rate, sensitivity or true positive rate, FPR, and AUC, mathematically represented as eqs. \eqref{Accuracy}-\eqref{AUC} respectively. 

\begin{equation}
Accuracy=\frac{TP + TN}{TP + TN + FP + FN}
\label{Accuracy}
\end{equation}

\begin{equation}
Specificity = \frac{TN}{TN+FP}
\label{Specificity}
\end{equation}

\begin{equation}
Senstivity = \frac{TP}{TP+FN}
\label{Senstivity}
\end{equation}

\begin{equation}
FPR = \frac{FP}{TN+FP}
\label{FPR}
\end{equation}

\begin{equation}
AUC = \int_{0}^{1} \frac{TP}{TP + FN}d\frac{FP}{FP + TN}
\label{AUC}
\end{equation}
Where true positive (TP) represents the number of correctly classified attack instances, true negative (TN) represents the number of correctly classified normal instances, false positive (FP) is the number of wrongly classified attack instances, and false negative (FN) is the number of wrongly classified normal instances. Accuracy is defined as the total number of correctly classified instances over the total number of instances in the dataset. Specificity is defined as the number of correctly classified normal instances over the total number of normal instances. Sensitivity is defined as the number of correctly classified attack instances over the total number of attack instances. FPR is defined as the number of incorrectly classified attack instances over the total number of normal instances. AUC refers to the area under the receiver operating characteristic (ROC) curve, where ROC curve defined as plotting TPR against FPR. 

\textcolor{black}{In order to perform a comprehensive performance assessment of different classifiers, we conducted experiments by using repeated hold-out as well as repeated $ k $-fold cross validation (10f) method \cite{kim2009estimating}. As suggested in \cite{repeatedKFold}, the repeated version stabilizes the error estimation and minimizes the variance of the validation approach.} For hold-out validation, we divided sample dataset into 60:40 ratio (60\% training instances and 40\% testing instances) in order to create train and test set. Similarly, for $ k $-fold cross validation the value of $ k $ is considered as 10. We considered 100 rounds of repeated 10f and hold-out validation as the classification models are observed to be stable, i.e., same prediction for the same test data. 10f is performed in order to asses the classifier's performance while avoiding the effect of instance sampling (i.e in case of hold-out validation).
\textcolor{black}{In order to avoid bias, all the performance results reported in this paper are the mean value of outputs from 10 iterations of each repeated validation approach. Each experiment is repeated by using different seed (an input to a random number generator) for avoiding biased results.}
\subsection{Statistical significance tests}
\textcolor{black}{In ML studies, comparison of multiple algorithms over multiple datasets is an essential issue \cite{demvsar2006statistical}. An algorithm may show better performance over one dataset whereas may fail to achieve similar result over another dataset. The reason for this may be the presence of outliers, feature distribution or algorithm characteristics. Thus, it becomes quite difficult to compare different algorithms among themselves. This consequently makes it challenging to decide which algorithm is better than others. To address this issue, the statistical assessment is needed to statistically validate the performance results.} 
\textcolor{black}{In this study, two statistical significance tests \cite{conover1980practical} are utilized in order to perform the comparison of classifiers in a correct way. Friedman \cite{friedman1937use} and Nemenyi \cite{dunn1961multiple} tests are selected for this purpose. The significance tests help in finding whether the classifiers are significantly different from each other or not \cite{demvsar2006statistical, garcia2008extension}. The null hypothesis ($ H_0 $) considered in this case is that there is no performance difference among classifiers. While alternative hypothesis ($ H_1 $) is that there is at least one classifier that performs significantly different that at least one other classifier. The main reason behind choosing the Friedman test is that it is the most powerful statistical test in case the number of entities being compared are greater than five \cite{conover1980practical,tama2019depth}. Friedman test helps in determining whether at least one classifier performs significantly better than the others in case of all the datasets. In any one such classifier is found, then Nemenyi post-hoc test is performed for pairwise multiple comparisons. As suggested in \cite{demvsar2006statistical}, it is crucial to conduct post-hoc test so as to identify the performance differences among the classifiers. 
	%Nemenyi test is chosen because it is one of the best-suited post-hoc test available for the Friedman test. 
	To be more specific, Friedman test checks for the significant difference among the classifiers being tested, whereas the Nemenyi test pinpoints where that difference lies. The further discussion assumes $ d $ as the number of datasets, and $ k $ as the number of classifiers. In Friedman test, initially the performance results ($ X_{ij} $) of classifiers are ranked ($ R(X_{ij}) $) for all the datasets. Then, sum of the $ R(X_{ij}) $ is computed for each classifier in order to obtain $ R_j $ (eq. \eqref{Rj}), where $ j = 1, 2, \ldots, k $. The Friedman statistic (F-Statistic) is computed as eq. \eqref{T2}, where $ Q $ is calculated as eq. \eqref{Q}.      }

\begin{equation}
\label{Rj}
R_j = \sum_{i = 1}^{d} R (X_{ij})
\end{equation}              

\begin{equation}
\label{T2}
F-Statistic  = \frac{(d - 1)Q}{d(k - 1) - Q}
\end{equation}

\begin{equation}
\label{Q}
Q = \frac{12}{dk(k + 1)} \sum_{j = 1}^{k} \Bigg( R_j - \frac{d(k + 1)}{2} \Bigg)^2
\end{equation}

The F-Statistic is tested against the F-quantiles for a given $ \alpha $  with degree of freedom, $ f_1 = k-1 $ and $f_2 = (d-1)(k-1)$, where $\alpha$ is the significance level being considered. In this study the values of $ d $, $ k $ are 4 and 7 respectively. Nemenyi post-hoc test is performed by calculating test statistic $ \gamma_{xy} $ (represented as eq. \eqref{Nemenyi}) for all classifier pairs, where $ \overline{R_x} $ and $ \overline{R_y} $ are mean ranks of classifiers $ x $ and $ y $ respectively on all datasets, and computed as eq. \eqref{R_j_bar}. After all the $ \gamma_{xy} $ are calculated and those which exceed a critical value are said to indicate a significant difference between classifiers $ x $ and $ y $ at $\alpha$ significance level. In this study, two values of $\alpha$ are considered, i.e., 0.05 and 0.1. The statistical analysis of both hold-out and 10f validation results is carried out in this experimental study.      

\begin{equation}
\label{Nemenyi}
\gamma_{xy} = \frac{\overline{R_x} - \overline{R_y}}{\sqrt{\frac{k(k+1)}{6d}}}
\end{equation}

\begin{equation}
\label{R_j_bar}
\overline{R_j} = \frac{1}{d} \sum_{i = 1}^{d} R (X_{ij})
\end{equation}

\section{Results and Analysis}\label{ResultsandAnalysis}
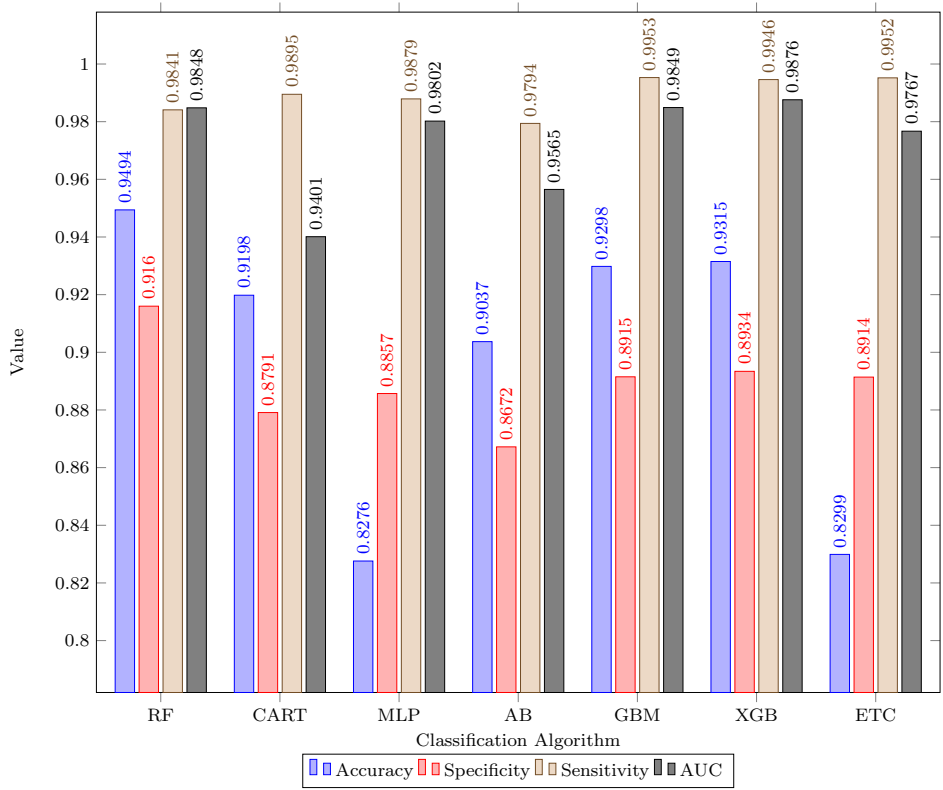
\begin{figure}[!h]
	\centering
	\begin{tikzpicture}[scale = .85]
	\begin{axis}[
	ybar,
	bar width=.3cm,
	every node near coord/.append style={rotate=90, anchor=west},
	width=1.2\textwidth,
	height=\textwidth,
	enlargelimits=0.09,
	legend style={at={(0.5,-.09)},
		anchor=north,legend columns=-1},
	ylabel={Value},
	xlabel={Classification Algorithm},
	symbolic x coords={RF, CART, MLP, AB, GBM, XGB, ETC},
	xtick=data,
	nodes near coords,
	every node near coord/.append style={/pgf/number format/.cd, fixed,precision=4},
	nodes near coords align={vertical},
	ymin=0.8,ymax=1.0,
	]
	
	\addplot coordinates {(RF, 0.9494
		) (CART,0.9198
		) (MLP,0.8276
		) (AB, 0.9037
		) (GBM,0.9298
		) (XGB,0.9315
		) (ETC,0.8299
		)};
	\addplot coordinates {(RF, 0.9160
		) (CART,0.8791
		) (MLP,0.8857
		) (AB,0.8672
		) (GBM,0.8915
		) (XGB, 0.8934
		) (ETC,0.8914
		)};
	\addplot coordinates {(RF, 0.9841
		) (CART,0.9895
		) (MLP,0.9879
		) (AB,0.9794
		) (GBM, 0.9953
		) (XGB,0.9946
		) (ETC, 0.9952
		)};
	\addplot coordinates {(RF,0.9848
		) (CART,0.9401
		) (MLP,0.9802
		) (AB,0.9565
		) (GBM, 0.9849
		) (XGB, 0.9876
		) (ETC,0.9767
		)};
	
	\legend{Accuracy, Specificity, Sensitivity, AUC}
	\end{axis}
	\end{tikzpicture}
	
	\caption{The average value of prominent metrics per classifier across four datasets with hold-out validation}
	\label{Fig1}
\end{figure}

\begin{figure}[!h]
	\centering
	\begin{tikzpicture}[scale = .85]
	\begin{axis}[
	ybar,
	bar width=.8cm,
	every node near coord/.append style={rotate=90, anchor=west},
	%	width=1.2\textwidth,
	%	height=.9\textwidth,
	enlargelimits=0.09,
	legend style={at={(.8,.95)},
		anchor=north,legend columns=-1},
	ylabel={Value},
	xlabel={Classification Algorithm},
	symbolic x coords={RF, CART, MLP, AB, GBM, XGB, ETC},
	xtick=data,
	nodes near coords,
	every node near coord/.append style={/pgf/number format/.cd, fixed,precision=4},
	nodes near coords align={vertical},
	%	every node near coord/.style={/pgf/number format/fixed},
	ymin=0,ymax=.4,
	]
	
	\addplot coordinates {(RF,0.0898
		) (CART,0.1207
		) (MLP,0.1141
		) (AB,0.1326
		) (GBM, 0.1084
		) (XGB, 0.1065
		) (ETC,0.1085
		)};
	
	\legend{FPR}
	\end{axis}
	\end{tikzpicture}
	
	\caption{The average value of FPR per classifier across four datasets with hold-out validation}
	\label{Fig2}
	
\end{figure}
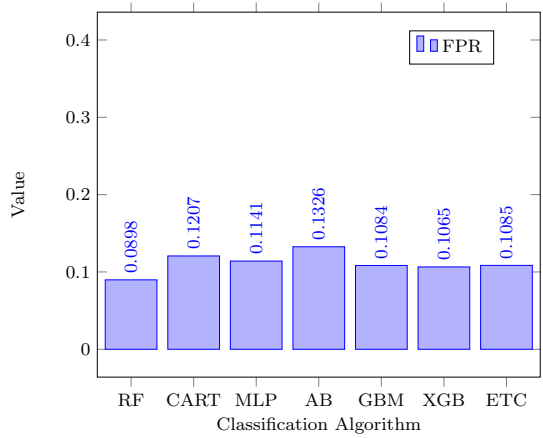

\textcolor{black}{In this section, a detailed discussion on performance analysis of ensembles (RF, AB, GBM, XGB and ETC) and single classifiers (CART and MLP) specific to CIDDS-001, UNSW-NB15, and NSL-KDD datasets is done. The  results are compared and statistically analyzed.} We have shown that the classifiers used in this study are suitable for intrusion detection in IoT applications. First, we analyze the performance results of hold-out validation. Fig. \ref{Fig1} indicates the average value of all prominent metrics other than FPR, achieved with hold-out validation across CIDDS-001, UNSW-NB15, KDDTrain+, and KDDTest+ datasets. It is observed that RF outperforms other classifiers in terms of accuracy (94.94\%) and specificity (91.6\%). GBM performs best in terms of sensitivity (99.53\%). In terms of AUC metric, XGB performs best by achieving 98.76\%. MLP is the worst performer in terms of accuracy (82.76\%), whereas AB performs worst in terms of specificity (86.72\%) and sensitivity (97.94\%). CART achieves lowest AUC value (94.01\%). Fig. \ref{Fig2} shows the average FPR values of classifiers across four datasets with hold-out validation. It is observed that RF performs best whereas AB performs worst among all the classifiers in terms of FPR by achieving 8.89\% and 13.26\% respectively. Table \ref{TableMBT} lists out model building time (MBT) of different classifiers across four datasets with hold-out validation. \textcolor{black}{The main reason behind computing MBT is that it is very important to consider the training time a model takes, as it would directly impact the resources usage, which is an important criterion for resource-constrained devices \cite{williams2006preliminary}.} Thus, MBT helps in making a good trade-off between resource usage and classification performance of a classifier, i.e., IDS. RF and CART take approximately 2 seconds for training on all four datasets. The highest time for model training is taken by GBM and ETC in case of KDDTrain+ dataset. MBT of all the classifiers is calculated for hold-out validation only.

\begin{table}[h]
	\centering
	\caption{MBT (seconds) of classifiers across four datasets}
	\label{TableMBT}
	\begin{tabular}{lrrrrrrr}
		\hline
		\textbf{Dataset}   & \textbf{RF} & \textbf{CART} & \textbf{MLP} & \textbf{AB} & \textbf{GBM} & \textbf{XGB} & \textbf{ETC} \\ \hline
		\textbf{CIDDS-001} & 0.4124      & 0.2353        & 1.0160       & 0.9557      & 20.1139      & 7.3965       & 17.5031      \\
		\textbf{UNSW-NB15} & 1.4657      & 0.6260        & 4.3782       & 7.9092      & 12.7477      & 23.7196      & 44.4775      \\
		\textbf{KDDTrain+} & 0.4087      & 0.2653        & 16.7050      & 2.6928      & 318.8914     & 14.0115      & 143.0728     \\ 
		\textbf{KDDTest+}  & 0.0601      & 0.0337        & 5.3062       & 0.4866      & 22.7327      & 2.9957       & 1.5041       \\ \hline
	\end{tabular}
	
\end{table}

Fig. \ref{Fig3} shows the average value of all prominent metrics other than FPR, achieved with 10f validation across CIDDS-001, UNSW-NB15, KDDTrain+, and KDDTest+ datasets. It is observed that the performances of all the used classifiers improve with 10f validation in comparison to classifier's performances with hold-out validation. This is due to the effect of sampling which results in the selection of random instances that leads to poor classification. This phenomenon advocates the use of 10f validation over hold-out validation. The 10f validation results show promising performance for all the classifiers. However, from the point of comparison, CART performs best in terms of accuracy (96.74\%). AB achieves the highest average value of specificity (97.5\%) metric. RF and XGB perform best in terms of sensitivity by achieving 97.31\% performance measure. For AUC, the best performing classifier is XGB which achieves 98.77\%. Fig. \ref{Fig4} shows the average FPR values of classifiers across four datasets with 10f validation. It is observed that CART performs best whereas RF performs worst among all classifiers in terms of FPR by achieving 3.78\% and 21.85\% respectively.

\begin{figure}[!h]
	\centering
	\begin{tikzpicture}[scale = .85]
	\begin{axis}[
	ybar,
	bar width=.3cm,
	every node near coord/.append style={rotate=90, anchor=west},
	width=1.2\textwidth,
	height=\textwidth,
	enlargelimits=0.09,
	legend style={at={(0.5,-.09)},
		anchor=north,legend columns=-1},
	ylabel={Value},
	xlabel={Classification Algorithm},
	symbolic x coords={RF, CART, MLP, AB, GBM, XGB, ETC},
	xtick=data,
	nodes near coords,
	every node near coord/.append style={/pgf/number format/.cd, fixed,precision=4},
	nodes near coords align={vertical},
	ymin=0.9,ymax=1.0,
	]
	
	\addplot coordinates {(RF, 0.9673
		) (CART,0.9674
		) (MLP,0.9673
		) (AB, 0.9673
		) (GBM,0.9673
		) (XGB,0.9673
		) (ETC,0.9673
		)};
	\addplot coordinates {(RF, 0.9619
		) (CART,0.9620
		) (MLP,0.9620
		) (AB,0.9750
		) (GBM,0.9618
		) (XGB, 0.9618
		) (ETC,0.9618
		)};
	\addplot coordinates {(RF, 0.9731
		) (CART,0.9730
		) (MLP,0.9730
		) (AB,0.9730
		) (GBM, 0.9730
		) (XGB,0.9731
		) (ETC, 0.9730
		)};
	\addplot coordinates {(RF, 0.9853
		) (CART,0.9422
		) (MLP,0.9813
		) (AB,0.9578
		) (GBM, 0.9851
		) (XGB, 0.9877
		) (ETC,0.9768
		)};
	
	\legend{Accuracy, Specificity, Sensitivity, AUC}
	\end{axis}
	\end{tikzpicture}
	
	\caption{The average value of prominent metrics per classifier across four datasets with 10f validation}
	\label{Fig3}
\end{figure}
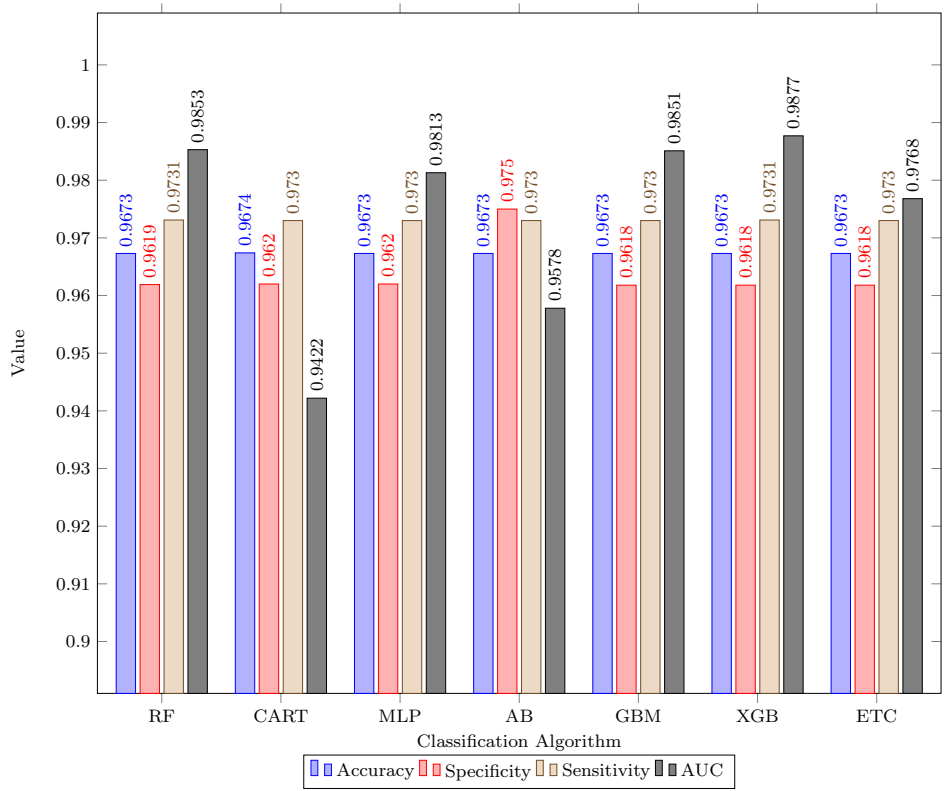

\begin{figure}[!h]
	\centering
	\begin{tikzpicture}[scale = .85]
	\begin{axis}[
	ybar,
	bar width=.8cm,
	every node near coord/.append style={rotate=90, anchor=west},
	%	width=1.2\textwidth,
	%	height=.9\textwidth,
	enlargelimits=0.09,
	legend style={at={(.8,.95)},
		anchor=north,legend columns=-1},
	ylabel={Value},
	xlabel={Classification Algorithm},
	symbolic x coords={RF, CART, MLP, AB, GBM, XGB, ETC},
	xtick=data,
	nodes near coords,
	nodes near coords align={vertical},
	%	every node near coord/.style={/pgf/number format/fixed},
	every node near coord/.append style={/pgf/number format/.cd, fixed,precision=4},
	ymin=0,ymax=.4,
	]
	
	\addplot coordinates {(RF,0.2185
		) (CART,0.0378
		) (MLP,0.0379
		) (AB,0.0380
		) (GBM, 0.2184
		) (XGB, 0.0380
		) (ETC,0.0380
		)};
	
	\legend{FPR}
	\end{axis}
	\end{tikzpicture}
	
	\caption{The average value of FPR per classifier across four datasets with 10f validation}
	\label{Fig4}
\end{figure}
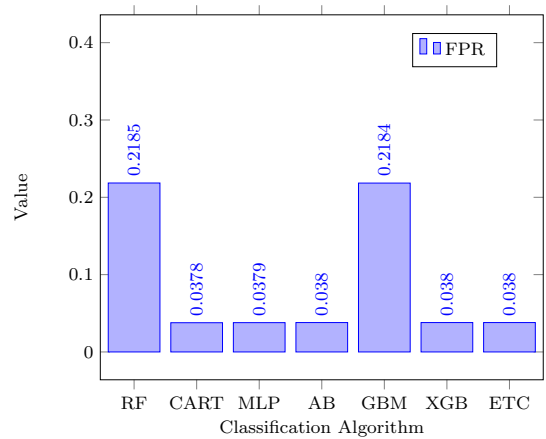

\begin{table}[!h]
	\centering
	\caption{Friedman test statistics for hold-out validation}
	\begin{tabular}{lrrrrr}
		\hline
		& Accuracy & Specificity & Sensitivity & FPR   & AUC       \\ \hline
		F-Statistic       & 6.7745   & 7.7091      & 7.1434     & 7.7091 & 4.7020  \\ 
		$ p $-value         & 0.0007   & 0.0003      & 0.0005     & 0.0003 & 0.0048  \\
		$ \alpha $ = 0.05 & R        & R           & R          & R      & R            \\ 
		$ \alpha $ = 0.1  & R        & R           & R          & R      & R           \\ \hline
	\end{tabular}
	
	\label{table1}
\end{table}

\begin{table}[!h]
	\centering
	\caption{Friedman test (mean ranks for hold-out validation)}
	\begin{tabular}{llllll}
		\hline
		& Accuracy       & Specificity    & Sensitivity     & FPR     & AUC                    \\ \hline
		RF   & 4.875          & 5.750         & 2.875          & 2.250         & 3.750                   \\
		CART & 2.250         & 2.000          & 3.250          & 6.000          & 1.500  \\
		MLP  & 3.250         & 3.250          & 2.250         & 4.750         & 3.000                   \\
		AB   & 1.250 & 1.500 & 2.000 & 6.500          & 2.875                 \\
		XGB  & 6.000          & 6.375          & 5.500         & 1.625 & 6.000               \\
		GBM  & 5.750         & 4.625          & 6.250         & 3.375          & 5.875                \\
		ETC  & 4.625          & 4.500          & 5.875          & 3.500         & 5.000                 \\ \hline
	\end{tabular}
	
	\label{table2}
\end{table}

The performance results are statistically assessed using Friedman and Nemenyi post-hoc test. Both the statistical tests are performed for two values of significance level $ \alpha $ (i.e., 0.05 and 0.1). For $ \alpha = \{0.05, 0.1\}$ the value of $ f_1 $, $ f_2 $ are 6 and 18 respectively, the F-Statistic and $ p $-value for each performance metric is computed. Table \ref{table1} shows Friedman test statistics for hold-out validation results. From the results it is observed that the performance of the classifiers is significantly different ($ p<0.05 $ and $ p<0.1 $) in terms of all the considered performance metrics. Thus, it is concluded that there is at least one classifier that performs significantly different that one another classifier. Because the results of Friedman test are highly significant ($ p<0.05 $ and $ p<0.1 $), the null hypothesis $ H_0 $ is rejected (represented by R in Table \ref{table1}) and alternative hypothesis $ H_1 $ is accepted. In Table \ref{table2}, the mean ranks of all the classifiers for hold-out validation are shown. In order to find which classifier pairs perform significantly different, Nemenyi post-hoc test is performed. For this purpose, the $ p $-value of all the pairwise comparisons is tested against the considered significance level $ \alpha $. 

Table \ref{table3} presents the results of the Nemenyi test (pairwise comparison) over accuracy, specificity and sensitivity. As shown in table \ref{table3}, the classifier's accuracy is highly significant ($ p<0.05 $) in case of AB-XGB pair, whereas less significant ($ p<0.1 $) in case of AB-GBM pair. While the remaining pairs are not significant ($ p<0.1 $). Moreover, in terms of specificity, the highly significant pair is AB-XGB, whereas the less significant pair is XGB-CART, whilst no other pair is found to be significant. Furthermore, no pair is significant in terms of sensitivity. Table \ref{table4} shows the results of the Nemenyi test (pairwise comparison) over FPR, AUC and MBT. It is observed that for FPR metric the classifier's performance is highly significant in the case of AB-XGB and less significant in the case of XGB-CART, whereas remaining pairs are not significant. While, in the case of AUC metric there are no pairs which are highly significant, whilst GBM-CART and XGB-CART are the only less significant pairs among all the classifier pairs, every other pair is not significant. 

Table \ref{table5} shows Friedman test statistics for 10f validation results. From the results it is observed that the performance of the classifiers is significantly different ($ p<0.05 $ and $ p<0.1 $) in terms of AUC only. Thus, it is concluded that there is at least one classifier that performs significantly different than one another classifier. Because the result of Friedman test is highly significant ($ p<0.05 $ and $ p<0.1 $) for AUC metrics, the null hypothesis $ H_0 $ is rejected and alternative hypothesis $ H_1 $ is accepted (represented by A in Table \ref{table5}). In Table \ref{table6} the mean ranks of all the classifiers for 10f validation results are shown. Table \ref{table7} presents the results of the Nemenyi test (pairwise comparison) over AUC values. As shown in table \ref{table7}, the classifier's AUC measure is found less significant in case of XGB-CART pair, whereas all other pairs are not significant.

%\begin{landscape}
\begin{table}[!h]
	\centering
	\bigbreak
	\bigbreak
	\bigbreak
	\bigbreak
	\caption{Nemenyi pairwise comparison (hold-out validation) Part I}
%	\resizebox{\textwidth}{!}{%
		\begin{tabular}{lllllllllllll}
			\hline
			\multicolumn{1}{c}{\multirow{3}{*}{$ A_1 ~versus~ A_2 $}} & \multicolumn{4}{c}{Accuracy}                   & \multicolumn{4}{c}{Specificity}                & \multicolumn{4}{c}{Senstivity}                 \\ \cline{2-13} 
			\multicolumn{1}{c}{}                          &           &         & \multicolumn{2}{c}{$ \alpha $} &           &         & \multicolumn{2}{c}{$ \alpha $} &           &         & \multicolumn{2}{c}{$ \alpha $} \\ \cline{4-5} \cline{8-9} \cline{12-13} 
			\multicolumn{1}{c}{}                          & F-Statistic & $ p $-value & 0.05        & 0.1        & F-Statistic & $ p $-value & 0.05        & 0.1        & F-Statistic & $ p $-value & 0.05        & 0.1        \\ \hline
			AB   versus XGB  & 3.1096 & 0.0393 & R & R & 3.1914 & 0.0297 & R & R & 2.2912 & 0.4608 & A & A \\
			AB   versus GBM  & 2.9459 & 0.0676 & A & R & 2.0457 & 0.8563 & A & A & 2.7822 & 0.1133 & A & A \\
			AB   versus RF   & 2.3731 & 0.3704 & A & A & 2.7822 & 0.1134 & A & A & 0.5728 &1.0     & A & A \\
			AB   versus ETC  & 2.2094 & 0.57   & A & A & 1.9639 &1.0     & A & A & 2.5367 & 0.2349 & A & A \\
			AB   versus MLP  & 1.3093 &1.0     & A & A & 1.1456 &1.0     & A & A & 0.4091 &1.0     & A & A \\
			AB   versus CART & 0.6546 &1.0     & A & A & 0.3273 &1.0     & A & A & 0.8183 &1.0     & A & A \\
			CART versus ETC  & 1.5548 &1.0     & A & A & 1.6366 &1.0     & A & A & 1.7184 &1.0     & A & A \\
			CART versus MLP  & 0.6546 &1.0     & A & A & 0.8183 &1.0     & A & A & 0.6546 &1.0     & A & A \\
			ETC  versus MLP  & 0.9001 &1.0     & A & A & 0.8183 &1.0     & A & A & 2.3731 & 0.3704 & A & A \\
			GBM  versus CART & 2.2912 & 0.4608 & A & A & 1.7184 &1.0     & A & A & 1.9639 &1.0     & A & A \\
			GBM  versus MLP  & 1.6366 &1.0     & A & A & 0.9001 &1.0     & A & A & 2.6186 & 0.1854 & A & A \\
			GBM  versus ETC  & 0.7364 &1.0     & A & A & 0.0818 &1.0     & A & A & 0.2455 &1.0     & A & A \\
			GBM  versus RF   & 0.5728 &1.0     & A & A & 0.7364 &1.0     & A & A & 2.2094 & 0.57   & A & A \\
			GBM  versus XGB  & 0.1636 &1.0     & A & A & 1.1456 &1.0     & A & A & 0.4909 &1.0     & A & A \\
			RF   versus CART & 1.7184 &1.0     & A & A & 2.4549 & 0.2959 & A & A & 0.2455 &1.0     & A & A \\
			RF   versus MLP  & 1.0638 &1.0     & A & A & 1.6366 &1.0     & A & A & 0.1636 &1.0     & A & A \\
			RF   versus ETC  & 0.1636 &1.0     & A & A & 0.8183 &1.0     & A & A & 1.9639 &1.0     & A & A \\
			XGB  versus CART & 2.4549 & 0.2959 & A & A & 2.8641 & 0.0878 & A & R & 1.4729 &1.0     & A & A \\
			XGB  versus MLP  & 1.8003 &1.0     & A & A & 2.0457 & 0.8563 & A & A & 2.1276 & 0.7007 & A & A \\
			XGB  versus ETC  & 0.9001 &1.0     & A & A & 1.2274 &1.0     & A & A & 0.2455 &1.0     & A & A \\
			XGB  versus RF   & 0.7364 &1.0     & A & A & 0.4091 &1.0     & A & A & 1.7184 &1.0     & A & A \\ \hline        
		\end{tabular}
		%
%	}
	\label{table3}
\end{table}
%\end{landscape}

\begin{table}[!h]
	\centering
	\caption{Nemenyi pairwise comparison (hold-out validation) Part II}
	%	\resizebox{\textwidth}{!}{%
	\begin{tabular}{lllllllll}
		\hline
		\multicolumn{1}{c}{\multirow{3}{*}{$ A_1 ~versus~ A_2 $}} & \multicolumn{4}{c}{FPR}                        & \multicolumn{4}{c}{AUC}                                            \\ \cline{2-9} 
		\multicolumn{1}{c}{}                          &           &         & \multicolumn{2}{c}{$ \alpha $} &           &     & \multicolumn{2}{c}{$ \alpha $}                    \\ \cline{4-5} \cline{8-9}  
		& F-Statistic & $ p $-value & 0.05        & 0.1        & F-Statistic & $ p $-value & 0.05      & 0.1              \\ \hline
		AB   versus XGB  & 3.1914 & 0.0297 & R  & R & 2.0457 & 0.8563 & A & A  \\
		AB   versus GBM  & 2.0457 & 0.8563 & A  & A & 1.9639 &1.0     & A & A \\
		AB   versus RF   & 2.7822 & 0.1133 & A  & A & 0.5728 &1.0     & A & A  \\
		AB   versus ETC  & 1.9639 &1.0     & A  & A & 1.3911 &1.0     & A & A  \\
		AB   versus MLP  & 1.1456 &1.0     & A  & A & 0.0818 &1.0     & A & A \\
		AB   versus CART & 0.3273 &1.0     & A  & A & 0.9001 &1.0     & A & A  \\
		CART versus ETC  & 1.6366 &1.0     & A  & A & 2.2912 & 0.4608 & A & A  \\
		CART versus MLP  & 0.8183 &1.0     & A  & A & 0.9819 &1.0     & A & A  \\
		ETC  versus MLP  & 0.8183 &1.0     & A & A & 1.3093 &1.0     & A & A  \\
		GBM  versus CART & 1.7184 &1.0     & A  & A & 2.8641 & 0.0878 & A & R  \\
		GBM  versus MLP  & 0.9001 &1.0     & A  & A & 1.8821 &1.0     & A & A  \\
		GBM  versus ETC  & 0.0818 &1.0     & A  & A & 0.5728 &1.0     & A & A  \\
		GBM  versus RF   & 0.7364 &1.0     & A  & A & 1.3911 &1.0     & A & A  \\
		GBM  versus XGB  & 1.1456 &1.0     & A  & A & 0.0818 &1.0     & A & A  \\
		RF   versus CART & 2.4549 & 0.2959 & A  & A & 1.4729 &1.0     & A & A  \\
		RF   versus MLP  & 1.6366 &1.0     & A  & A & 0.4909 &1.0     & A & A  \\
		RF   versus ETC  & 0.8183 &1.0     & A  & A & 0.8183 &1.0     & A & A  \\
		XGB  versus CART & 2.8641 & 0.0878 & A  & R & 2.9459 & 0.0676 & A & R  \\
		XGB  versus MLP  & 2.0457 & 0.8563 & A  & A & 1.9639 &1.0     & A & A  \\
		XGB  versus ETC  & 1.2274 &1.0     & A  & A & 0.6546 &1.0     & A & A  \\
		XGB  versus RF   & 0.4091 &1.0     & A  & A & 1.4729 &1.0     & A & A \\ \hline       
	\end{tabular}
	%	%
	%	}
	\label{table4}
\end{table}

\begin{table}[!h]
	\centering
	\caption{Friedman test statistics for 10f validation}
	\begin{tabular}{llllll}
		\hline
		& Accuracy &Specificity & Sensitivity &FPR    & AUC        \\ \hline
		F-Statistic   & 0.1698   & 0.2346      & 0.2740      & 0.4242 & 4.5294  \\ 
		$ p $-value     & 0.9816   & 0.9594      & 0.9418     & 0.8532 & 0.0057   \\
		$ \alpha $ = 0.05 & A        & A           & A          & A      & R           \\ 
		$ \alpha $ = 0.1  & A        & A           & A          & A      & R          \\ \hline
	\end{tabular}
	
	\label{table5}
\end{table}

\begin{table}[!h]
	\centering
	\caption{Friedman test (mean ranks for 10f validation)}
	\begin{tabular}{llllll}
		\hline
		& Accuracy      & Specificity   & Sensitivity     & FPR           & AUC                      \\ \hline
		RF   & 4.000         & 4.000          & 4.500          & 4.625          & 3.625           \\ 
		CART & 4.375          & 4.250         & 3.375          & 3.500          & 1.500 \\
		MLP  & 4.250          & 5.000         & 3.375          & 2.750 & 3.125                  \\
		AB   & 4.625          & 4.000          & 4.500         & 4.750          & 2.875                 \\
		XGB  & 3.125 & 3.375          & 4.500          & 4.250          & 6.125                  \\
		GBM  & 4.000         & 3.250& 4.500          & 4.625          & 5.625                  \\ 
		ETC  & 3.625          & 4.125          & 3.250& 3.500          & 5.125                  \\ \hline
	\end{tabular}
	
	\label{table6}
\end{table}

\begin{table}[!h]
	\centering
	\caption{Nemenyi test (10f validation)}
	%		\resizebox{\textwidth}{!}{%
	\begin{tabular}{lllll}
		\hline
		\multicolumn{1}{c}{\multirow{3}{*}{$ A_1 ~versus~ A_2 $}} & \multicolumn{4}{c}{AUC}                                      \\ \cline{2-5} 
		&  & & \multicolumn{2}{c}{$ \alpha $}                                        \\ \cline{4-5}  
		& F-Statistic       & $ p $-value         & 0.05        & 0.1        \\ \hline
		AB   versus XGB  & 2.1276 & 0.7007 & A & A \\
		AB   versus GBM  & 1.8003 &1.0     & A & A  \\
		AB   versus ETC  & 1.4729 &1.0     & A & A \\
		AB   versus CART & 0.9001 &1.0     & A & A \\
		AB   versus RF   & 0.4909 &1.0     & A & A \\
		AB   versus MLP  & 0.1636 &1.0     & A & A\\
		CART versus GBM  & 2.7004 & 0.1454 & A & A \\
		CART versus ETC  & 2.3731 & 0.3704 & A & A \\
		CART versus MLP  & 1.0638 &1.0     & A & A   \\
		ETC  versus MLP  & 1.3093 &1.0     & A & A \\
		ETC  versus GBM  & 0.3273 &1.0     & A & A\\
		MLP  versus GBM  & 1.6366 &1.0     & A & A \\
		RF   versus CART & 1.3911 &1.0     & A & A  \\
		RF   versus GBM  & 1.3093 &1.0     & A & A  \\
		RF   versus ETC  & 0.9819 &1.0     & A & A \\
		RF   versus MLP  & 0.3273 &1.0     & A & A \\
		XGB  versus CART & 3.0277 & 0.0517 & A & R \\
		XGB  versus MLP  & 1.9639 &1.0     & A & A \\
		XGB  versus RF   & 1.6366 &1.0     & A & A\\
		XGB  versus ETC  & 0.6546 &1.0     & A & A  \\
		XGB  versus GBM  & 0.3273 &1.0     & A & A \\ \hline
	\end{tabular}
	%	%
	%	}
	\label{table7}	
\end{table}

In addition, we have analyzed the average response time (seconds) that a classifier takes to classify an instance. The main reason to perform this experiment is that the knowledge of classifier's response time plays an important role in its selection as a intrusion detection system \cite{hodo2016threat}. The classifiers with quick (small) response time are favored over classifier with slow (large) response time. To accomplish this task, all the classifiers with test data as input were executed on Raspberry Pi 3 Model B+. The average time is computed by dividing the total time taken by the classifier for classifying all the test instances by the total number of test instances. 

\begin{equation}
\label{classificationtime}
Average~response~time = \frac{\sum_{i = 1}^{n_{test}}t_i}{n_{test}}
\end{equation} 

Eq. \eqref{classificationtime} represents the mathematical expression of average response time, where $ i $ represents an instance number, $ t_i $ represents time taken by a classifier to classify $ i^{th} $ test instance into attack or normal category, and $ n_{test} $ is the total number of test instances. 

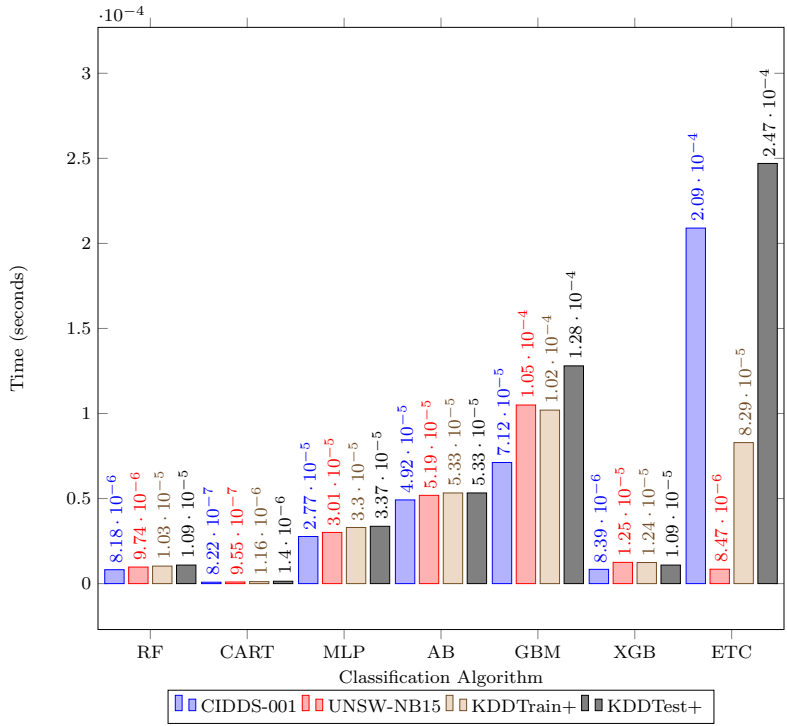
\begin{figure}[tbh]
	\centering
	\begin{tikzpicture}[scale = .85]
	\begin{axis}[
	ybar,
	bar width=.3cm,
	every node near coord/.append style={rotate=90, anchor=west},
	width=\textwidth,
	height=.9\textwidth,
	enlargelimits=0.09,
	legend style={at={(0.5,-.1)},
		anchor=north,legend columns=-1},
	ylabel={Time (seconds)},
	xlabel={Classification Algorithm},
	symbolic x coords={RF, CART, MLP, AB, GBM, XGB, ETC},
	xtick=data,
	nodes near coords,
	nodes near coords align={vertical},
	ymin=0,ymax=3E-04,
	]
	\addplot coordinates {(RF,8.18E-06
		) (CART,8.22E-07
		) (MLP,2.77E-05
		) (AB,4.92E-05
		) (GBM, 7.12E-05
		) (XGB, 8.39E-06
		) (ETC,2.09E-04
		)};
	\addplot coordinates {(RF,9.74E-06
		) (CART,9.55E-07
		) (MLP,3.01E-05
		) (AB,5.19E-05
		) (GBM, 1.05E-04
		) (XGB, 1.25E-05
		) (ETC,8.47E-06
		)};
	\addplot coordinates {(RF,1.03E-05
		) (CART,1.16E-06
		) (MLP,3.30E-05
		) (AB,5.33E-05
		) (GBM, 1.02E-04
		) (XGB, 1.24E-05
		) (ETC,8.29E-05
		)};
	\addplot coordinates {(RF,1.09E-05
		) (CART,1.40E-06
		) (MLP,3.37E-05
		) (AB,5.33E-05
		) (GBM, 1.28E-04
		) (XGB, 1.09E-05
		) (ETC, 2.47E-04
		)};
	
	\legend{CIDDS-001, UNSW-NB15, KDDTrain+, KDDTest+}
	\end{axis}
	\end{tikzpicture}
	
	\caption{Average response time of classifiers}
	\label{Fig5}
\end{figure}                              

Fig. \ref{Fig5} shows the average response time taken by different classifiers for classifying a single instance. From the experiment results, it is observed that CART takes minimum time to classify an instance of CIDDS-001, UNSW-NB15, KDDTrain+, and KDDTest+ in comparison to other classifiers. RF and XGB show almost similar results in terms of average response time for all four datasets. ETC takes maximum time for classifying an instance of CIDDS-001 and KDDTest+ dataset in comparison to other classifiers. Moreover, GBM is the worst performer in the case of KDDTrain+ dataset. The experimental results show promising solutions for the choice of different classifiers suitable for performing the task of intrusion detection (DoS specific) in IoT applications. The classifiers have been validated using hold-out and 10f validation methods. Both the methods show promising performance results in terms of accuracy, specificity, sensitivity, FPR, AUC. \textcolor{black}{These results can be used to select the suitable classifier as per the requirement of the application. Like, if an application demands high accuracy and low FPR, then CART, MLP, AB, XGB, or ETC can be used.} Whereas, if an application demands quick response time, then CART, RF, or XGB can be selected. Similar trade-offs can be considered in the selection of the best suitable classifier for an IoT application. The real-time performance of IDS depends on the dataset selected for model training. Thus, a dataset containing traffic patterns of recent types of DoS attacks must be used for achieving the best real-time results. \textcolor{black}{CIDDS-001 and UNSW-NB15 are suitable choices for this purpose. It can be observed from the experimental results that classifiers show promising performance results with CIDDS-001 and UNSW-NB15 dataset thus, we suggest these datasets for training IDSs to achieve the best classification results.}  

%\subsection{Significance of supervised learning} 
%\subsection{Impact of unsupervised learning}
\textcolor{black}{In this study only supervised learning based ML classifiers are used. It is one of the popular ML approaches in which the classifier uses known target values for training. The result shown by different ML classifiers shows the effectiveness of using supervised learning for the intrusion detection task. The main reason for choosing supervised learning is that the network characteristics (traffic patterns) can be effectively used to train ML models for further predictions. These patterns can be differentiated between normal and attack based on different network based features. The other ML approach like unsupervised learning can also be used to perform a similar task. Clustering (i.e., unsupervised learning algorithm) can be used to train the different classifiers. In this paper, the emphasis is made particularly on the performance assessment of supervised ML algorithms. The performance assessment of unsupervised ML algorithms for intrusion detection in IoT will be considered in our future work.}  

\section{Conclusion}\label{Conclusion}
In this paper, a study on anomaly-based IDS suitable for securing IoT against DoS attacks is carried out. The performance assessment of seven machine learning classification algorithms including random forests, adaboost, gradient boosted machine, extremely randomized trees, classification and regression trees, and multi-layer perceptron is done. The optimal parameters of classifiers are obtained using a random search algorithm. Performance of all the classifiers is measured in terms of accuracy, specificity, sensitivity, false positive rate, and area under the receiver operating characteristic curve. Benchmarking of all the classifiers is performed on CIDDS-001, UNSW-NB15, and NSL-KDD datasets. Moreover, in order to find significant differences among classifiers the statistical analysis of performance measures is done using Friedman and Nemenyi post host tests. In addition to this, the average response time of all classifiers is evaluated on Raspberry Pi hardware device. \textcolor{black}{From the performance results and statistical tests, it is concluded that classification and regression trees, and extreme gradient boosting classifier show the best trade-off between prominent metrics and response time, thus both are the suitable choice for building IoT specific anomaly-based IDS.} Our future goal is to design an IDS for defending routing attacks in IoT networks.

\section*{ACKNOWLEDGMENT}
This research was supported by the Ministry of Human Resource Development, Government of India.

\section*{Conflict of Interest}
On behalf of all authors, the corresponding author states that there is no conflict of interest.

%\section*{References}
\bibliographystyle{spbasic}   
%\bibliography{mybibfile}

%\clearpage

\end{document}